\newif\iflandscape
\newif\ifportrait
\typein[\lorp]%
{Typein "l" (for landscape,) or "p" (for portrait, onecolumn)}
\if l\lorp \landscapetrue \else \portraittrue \fi
\newlength{\extralineskip}
\ifportrait
  \documentstyle[epsf]{article}
\fi
\iflandscape
  \documentstyle[twocolumn,epsf]{article}
\fi

\def\bea{\begin{eqnarray}}
\def\eea{\end{eqnarray}}
\def\nn{\nonumber}

\def\beq{\begin{equation}}
\def\eeq{\end{equation}}
\def\ba{\beq\new\begin{array}{c}}
\def\ea{\end{array}\eeq}
\def\be{\ba}
\def\ee{\ea}

\makeatletter
\newdimen\normalarrayskip              % skip between lines
\newdimen\minarrayskip                 % minimal skip between lines
\normalarrayskip\baselineskip
\minarrayskip\jot
\newif\ifold             \oldtrue            \def\new{\oldfalse}
\def\arraymode{\ifold\relax\else\displaystyle\fi} % mode of array entries
\def\eqnumphantom{\phantom{(\theequation)}}     % right phantom in eqnarray
\def\@arrayskip{\ifold\baselineskip\z@\lineskip\z@
     \else
     \baselineskip\minarrayskip\lineskip2\minarrayskip\fi}
\def\@arrayclassz{\ifcase \@lastchclass \@acolampacol \or
\@ampacol \or \or \or \@addamp \or
   \@acolampacol \or \@firstampfalse \@acol \fi
\edef\@preamble{\@preamble
  \ifcase \@chnum
     \hfil$\relax\arraymode\@sharp$\hfil
     \or $\relax\arraymode\@sharp$\hfil
     \or \hfil$\relax\arraymode\@sharp$\fi}}
\def\@array[#1]#2{\setbox\@arstrutbox=\hbox{\vrule
     height\arraystretch \ht\strutbox
     depth\arraystretch \dp\strutbox
     width\z@}\@mkpream{#2}\edef\@preamble{\halign \noexpand\@halignto
\bgroup \tabskip\z@ \@arstrut \@preamble \tabskip\z@ \cr}%
\let\@startpbox\@@startpbox \let\@endpbox\@@endpbox
  \if #1t\vtop \else \if#1b\vbox \else \vcenter \fi\fi
  \bgroup \let\par\relax
  \let\@sharp##\let\protect\relax
  \@arrayskip\@preamble}
\def\eqnarray{\stepcounter{equation}%
              \let\@currentlabel=\theequation
              \global\@eqnswtrue
              \global\@eqcnt\z@
              \tabskip\@centering
              \let\\=\@eqncr
              $$%
 \halign to \displaywidth\bgroup
    \eqnumphantom\@eqnsel\hskip\@centering
    $\displaystyle \tabskip\z@ {##}$%
    Üglobal\@eqcnt\@ne \hskip 2\arraycolsep
         $\displaystyle\arraymode{##}$\hfil
    Üglobal\@eqcnt\tw@ \hskip 2\arraycolsep
         $\displaystyle\tabskip\z@{##}$\hfil
         \tabskip\@centering
    &{##}\tabskip\z@\cr}
\newfont{\hr}{msbm10}
\newfont{\ams}{msam10}
\begin{document}
\begin{titlepage}
\setcounter{footnote}0
\begin{center}
\hfill UUITP-6/95, ITEP-M3/95, FIAN/TD-9/95\\
\vspace{0.3in}
\ifportrait
{\LARGE\bf Integrability and Seiberg-Witten Exact Solution}
\fi
\iflandscape
{\LARGE\bf Integrability and Seiberg-Witten Exact Solution}
\fi
\\[.2in]
{\Large A.Gorsky
\footnote{E-mail address: gorsky@butp.unibe.ch, sasha@rhea.teorfys.uu.se}
$^{\dag}$}\\
{\it Institute for Theoretical Physics, Bern University,
Sidlerstrasse 5, Bern, Switzerland}\\
\bigskip
{\Large I.Krichever
\footnote{E-mail address: krichev@sfb288.math.tu-berlin.de}}\\
{\it L.D.Landau Institute for Theoretical Physics, 117940 Moscow, Russia}\\
\bigskip
{\Large A.Marshakov
\footnote{E-mail address:
mars@lpi.ac.ru, andrei@rhea.teorfys.uu.se, marshakov@nbivax.nbi.dk}
$^{\ddag}$,
A.Mironov
\footnote{E-mail address:
mironov@lpi.ac.ru, mironov@grotte.teorfys.uu.se}$^{\ddag}$},\\
{\it Institute of Theoretical Physics, Uppsala University,
Uppsala S-75121,
Sweden}\\
\bigskip
{\Large A.Morozov
\footnote{E-mail address:
morozov@vxdesy.desy.de}}\\
{\it ITEP, Moscow, 117 259, Russia}\\
\bigskip
$\phantom{gh}^{\dag}$Permanent address:  {\it ITEP, Moscow,
117 259, Russia}\\
$\phantom{gh}^{\ddag}$Permanent address:  {\it Theory Department,
P. N. Lebedev Physics
Institute , Leninsky prospect, 53, Moscow,~117924, Russia\\
and ITEP, Moscow 117259, Russia}
\end{center}

\bigskip

\begin{quotation}
The exact  Seiberg-Witten (SW) description of the light sector in the
$N=2$ SUSY $4d$
Yang-Mills theory \cite{SW} is reformulated in terms of
integrable systems and appears to be a
Gurevich-Pitaevsky (GP) \cite{GP} solution to the elliptic Whitham
equations. We consider this as an implication that dynamical
mechanism behind the SW solution is related to integrable
systems on the moduli space of instantons. We emphasize the role
of the Whitham theory as a possible substitute of
the renormalization-group approach to the construction of low-energy
effective actions.
\end{quotation}
\end{titlepage}
\clearpage
\newpage

The exact expression for the vacuum-condensate
dependence of effective coupling constant in $d=4$ $N=2$ SUSY YM
theory \cite{SW} provides a new basis for the search of a relevant
description
of vacuum structure in non-abelian theories. Especially
interesting is
emergence of characteristic features of $2d$-integrable structures
in essentially $4d$ problem.
In this letter we explain that the SW
{\it answer} for $4d$ theory is just the same
as the GP solution of elliptic Whitham equations, which in its
turn is a simple analog of solutions to
``string equations'' arising in the context of $2d$ (world sheet)
string theories and gravity models
\footnote{To be exact we discuss throughout this letter the {\it
first} GP solution, which arises as a step decay in the KdV theory,
while the {\it second} one rather corresponds to $2d$ string
equations.}.
A more detailed discussion will be presented elsewhere.

\bigskip

1.  We begin with a survey of the relevant statements from the
general theory of  $4d$ YM fields and from ref.\cite{SW}.

The simplest dynamical characteristic of YM theory is the effective
coupling constant
$g^{-2}(\mu)$ (defined as a coefficient in front of  $\int{\rm
tr} F_{\mu\nu}^2 $ in effective action)
as a function of normalization point $\mu$ (roughly speaking,
the IR cut-off in the integration over fast quantum fluctuations).

Within the perturbation theory for non-abelian model this function
is given by Fig.1a. If there is a scalar-field condensate
\footnote{See \cite{SW} for notational details.} $u$,
spontaneously breaking original gauge symmetry $SU(2)$ to $U(1)$,
one gets instead the picture like Fig.1b. If the original symmetry
is larger than $SU(2)$ there is a series of transitions at
various points $u_a$. In the $N=2$ SUSY YM theory the $U(1)$
$\beta$-function  is zero, and what one obtains is Fig.1c. In this
model one is actually interested in the function $g^{-2}(\mu | u)$ of
{\it two} variables, since there is a valley in the effective
potential
and the value of $u$ is {\it a priori} arbitrary ($u$ is a dynamical
variable). Because of the simple pattern in Fig.1c, the function
\be
g^{-2}(\mu=0 | u )  = g^{-2}(\mu= u | 0)
\label{1}
\ee
can be considered as carrying a certain information about the
most intriguing
quantity $g^{-2}(\mu | u=0)$. In other words, one can substitute the
typical confinement-phase problem (of evaluation of
$g^{-2}(\mu | 0)$) by the typical Higgs-phase one (of evaluation of
$g^{-2}(0 | u)$) and the latter one definitely makes sense even
beyond the perturbation theory. It is also natural to introduce the
full complex
coupling constant
$\tau = \frac{1}{2\pi}(ig^{-2} + \theta)$, where $\theta$ is the
coefficient in front of the "topological" term
$\int{\rm tr} F_{\mu\nu}\tilde F_{\mu\nu} $.
Within the perturbation theory  $\theta$ does not depend on $\mu$ and
$u$ (see, however, \cite{AI}).

The definition of $\tau(\mu)$, as well as identifications like
eq.(\ref{1}), beyond the perturbation theory gets ambiguous.
However, a qualitative description is well known in the instanton-gas
approximation \cite{CDG}.
The new behaiviour, as compared to the perturbation theory,
is the occurence of $\mu$-dependence of $\theta$, which results in
renormalization of the bare $\theta \neq \pi$ at $\mu = \infty$ to
$\theta = 0$ at $\mu = \Lambda$, and deconfinement
(occurence of zero of $\beta$-function at
$g^{-2} \neq 0$) at $\theta = \pi$. Both effects are described
\cite{KM}, \cite{GS}
by the characteristic renormalization-group flow shown in Fig.2a.
The analytic description is given by equations

\begin{equation}\label{knimo}
\frac{d g^{-2}}{d \log\mu} = b + c e^{-\gamma g^{-2}}\cos\theta ,
$$
$$
\frac{d\theta}{d\log\mu} = c e^{-\gamma g^{-2}}\sin\theta ,
\end{equation}
where  $b(g^2) = b_1 + o(g^2)$ and $c(g^2,\mu,u)$ are
some positive functions depending on a particular model.

Beyond the instanton-gas approximation one should represent
$\tau(\mu)$
as some parameter of the effective theory on the universal moduli
space of instantons.\footnote{In general this theory can have
different phases. One of them -- believed to be relevant for
confinement in QCD -- is known in less formal terms as that
of instanton fluid \cite{DPS}.}
In the $N=2$ SUSY case, where perturbation theory
is almost trivial (for example, the perturbative $\beta$-function has
only one-loop contributions, \cite{NSVZ}), one can expect that the
relevant dynamical system is especially simple. One of the possible
ideas  is that
it somehow possesses integrable properties, peculiar for dynamics
on known moduli spaces (see for example \cite{H}, \cite{FR},
\cite{GN} ). The results of \cite{SW}, as well as their
generalizations in \cite{gen}, look as being consistent with this
integrable dynamics.

Namely, in \cite{SW}  $\tau(\mu=0 | u)$ is identified with the
coordinate on the modular half-plane for the one-dimensional complex
tori,
see Fig.2b, while $u$ is interpreted as a parameter (one of
ramification points) in their elliptic
representation
\begin{equation}\label{ell}
y^2 = (z^2 - \Lambda^2)(z - u)
\end{equation}
 i.e.

\begin{equation}\label{modfun}
u = \Lambda\left(1 - 2\frac{\theta_2^4(0|\tau)}
{\theta_3^4(0|\tau)} \right)
\end{equation}
The $SU(N)$ generalizations  are described in \cite{gen} in terms of
moduli of the specific subclass of hyperelliptic surfaces,

\be
y^2 = P_N^2(z) - \Lambda^2,
\label{speccu}
\ee
where $P_N$ is any polynomial of degree $N$.
These expressions provide an explicit way to avoid the singular
point $u = \Lambda$ (where $\tau = 0$, i.e.
$g^{-2} = 0$ and $\theta = 0$ -- in accordance with qualitative
Fig.2a; note that $u$ was restricted to be real in that picture)
by analytic continuation into the complex $u$-plane.
It also introduces one more singularity at another
point $u = -\Lambda$ ($\tau = \pm 1$), while the
vicinity of the last singular point $u = \infty$ ($\tau = i\infty$)
is described by the ordinary perturbation theory (thus,
the three ``infinitely-remote'' points are not identical, and the
theory
actually lives on the covering of moduli space -- again, as suggested
by naive Fig.2a .

Most impressive, \cite{SW} implies that the Riemann surfaces
themselves
-- not just their moduli -- have some physical significance. Namely, the
spectrum of excitations in the theory is identified as

\be
M_{m,n} = \left| m a + n a_D \right|
\label{spec}
\ee
where
\be\label{aaD}
a = \oint_A \lambda, \  \ \ a_D = \oint_B \lambda
\ee
and

\be\label{qumech}
\lambda = \frac{z-u}{y(z)}dz = \sqrt{u - \Lambda \cos \varphi}
d\varphi
\label{lambda}
\ee
is a particular 1-differential {\it on the surface} with the double
pole and the double zero at the ramification points
$z = \infty$ and $z = u$ respectively.\footnote{\label{f23}
In terms of $\varphi$-parametrization of the spectral
curve (see (\ref{lambda})),
the integrals (\ref{aaD}) for $|u|<1$ are just the action-integrals
$\oint p(\varphi)d\varphi$ in the Sine-Gordon model
$\ \ L_{SG} = \dot \varphi^2 - \Lambda\cos \varphi\ \ $
over the classically allowed and forbidden domains at a given
"energy" $u$, see Fig.3. Note that in this parametrization
$\varphi$ is identified with $\varphi + 2\pi$.
It also deserves mentioning that
for the elliptic solution the surface itself is isomorphic to its
Jacobian, thus the periods of differentials play the role of
periods of  the real motion in potential (\ref{qumech}) of
the ''auxuliary" quantum-mechanical problem.}

This poses the question of {\it what} is the {\it reason} for Riemann
surfaces to appear in this theory:
while $\tau$ and $u$ are present in it from the very beginning,
the surface is something new and emerges dynamically
only in the low-energy effective theory.

\bigskip

2. The answer to this question is, of course, more general
than the particular SW example
\footnote{The answer seems to be similar to that one from the $2d$
(string theory) case where the arising  (target-space!) spectral
curve might be associated with the ''scale-parameter" curve. The
non-perturbative effects imply that such a surface has a nontrivial
topology while the mechanism of arising the higher topologies is
not yet clear.}.

The effective dynamics in the space of coupling constants,
like $\theta$ and $g^{-2}$, substitutes original dynamics
in the ordinary space-time by a set of Ward identities
(low-energy theorems), which normally have the form
of non-linear differential equations for effective action
(which in this context is often refered to as generalized
$\tau$-function). When these equations belong to
(generalization of) KP/Toda-type hierarchy -- as it often happens
after appropriate choice of variables -- their solutions (i.e.
acceptable shapes of effective actions) are parametrized
in terms of some auxiliary ``spectral surfaces''
also known as ``target space'' curves (not world-sheet) in the
language of
string theory.

The family of ``vacua'' of the original model is thus naturally
associated with the family of spectral surfaces, i.e. with
their moduli space. It seems that only the moduli space
itself has physical meaning, not the spectral surfaces but
this is not however quite true.
So far we discussed the effective action (KP/Toda-like $\tau$-function)
as a function of time-variables (coupling constants
$\theta$, $g^{-2}$ etc). However, if considered as a function of
{\it moduli} (e.g. of scalar condensates $u$), the {\it effective}
$\tau$-function induces a new (low-energy-sector) dynamics
on the space of moduli. This new dynamics implies that the moduli are
no longer invariants of motion: instead they are ``RG-slow''
dynamical variables of the theory\footnote{The situation is much
similar to the standard renormalization group. Indeed, the RG dynamics
is governed by the action of some vector field $ {d \over \log{\mu}} =
\sum{\beta _i(g){\partial \over \partial g_i}}$ on the space of
coupling constants. Analogously, the Whitham dynamics gives an example
of {\it some} vector fields generated by the corresponding "slow"-time
flows, the counterpart of the coupling constant space being the moduli
space. Certainly, the standard RG approach is unambigously used only
within the perturbative framework, while we deal with the exact solution.
Therefore, we consider Whitham as the corresponding generalization of
the RG equations beyond the perturbative regime, which would still
have the form of the first-order differential equations with respect
to the coupling constants (co-ordinates in the moduli space).} . The
general approach to
construction of such effective actions is known as Bogolyubov-Whitham
averaging method (see \cite{DN}, \cite{Kr2} for a comprehensive review
and
references). Though this Whitham dynamics is that of the moduli,
its explicit formulation is most simple and natural in terms of
connections on spectral surfaces. Thus low-energy dynamics
actually gives a lot of physical significance to the spectral
surfaces themselves, and, after all, it is not such a big
surprise that dynamical characteristics -- of which (\ref{spec})
is a simplest example -- are expressed in terms of them.

In the SW case one could try to be more specific -- but in this letter
we restrict ourselves to the following simplified scheme.

One begins with considering the $u$-field-dependent dynamics on
the moduli space of instantons.
One can further think that some directions in the functional space
are most important for the low-energy theory. An obvious
candidate for such variable is
$K = \int \epsilon_{ijk}(A_i\partial_j A_k + \frac{2}{3}
A_iA_jA_k)d^3\vec r$.
Effective potentials are periodic in $K$ and associated excitations
are always light (unless they {\it mix} with something else which is
also light -- as is the case with the $\eta'$-meson in QCD).
The conjugate variable to $K$ is exactly $\theta$: one of our
most significant (along with $g^{-2}$ ``time variables'').
After ``Legendre transform'' one can think of original dynamics
of $u$-fields as of a RG-like one in the space of coupling
constants.  Solutions to these ``RG''-equations
identify the valley vacuum averages with moduli of spectral surfaces.
Monodromies on these surfaces are natural variables of the Toda-chain
hierarchies, with the length of the chain equal to $N$ for the
$SU(N)$ gauge group.
Indeed, the simplest type of dynamics for a variable $U$ in the
fundamental representation of $SU(N)$ is implied by Lagrangian
\be
{\cal L}_U = {\rm tr}(U^{-1}\partial U)^2 + {\rm tr} U,
\ee
which in the Cartan sector reduces to:
\be
{\cal L}_{eff} = \sum_{a =1}^{N} \left((\partial_{\mu}\phi_a)^2 +
e^{\phi_a}\right), \nn \\
U \longrightarrow {\rm diag}\left(e^{\phi_a}\right),
\ \ \ \sum_{a=1}^N \phi_a = 0.
\ee
Possible higher-derivative corrections to ${\cal L}_U$
can be associated with the higher Hamiltonians of the Toda-chain
hierarchy.

Now comes the first miracle. According to \cite{DuKri},
the finite-gap solutions to the Toda-chain systems are characterized
exactly by
hyperelliptic surfaces of the peculiar type (\ref{speccu}).\footnote{
We remind that the data (a spectral complex curve, a point on it
and a complex coordinate in the vicinity of the point)
is always in one-to-one correspondence with the solutions
to KP-hierarchy, explicit relation being given in terms of the
Baker-Ahiezer function (the curve itself can be also described by
the evolution-invariant equation
$\ {\rm det}\left( L(z) - y \right) = 0$,
where $L(z)$ is the Lax-operator). Particular
reductions of KP correspond to restrictions on the choice of
Riemann surfaces.
In particular, generic hyperelliptic surfaces correspond to
solutions to KdV,
while the subclass (\ref{speccu}) describes solutions to Toda-chain
hierarchy. The most spectacular in the last relation is that the
power of polynomial $P_N$ in (\ref{speccu}) is exactly the length of
the chain, i.e. the size of the $SU(N)$ matrices in the fundamental
representation.}

The next task is to consider effective Whitham dynamics.
With the ``first miracle'' in mind -- and with the
knowledge that all the Toda-chain systems are particular members
of the KP/Toda-lattice family -- we can just use the well-known
Whitham theory of {\it integrable} hierarchies \cite{DN}, \cite{Kr2}
\footnote{
This context can actually be not so arrow as it seems. As often
happens, different original (non-renormalized) models produce
the same kind of effective (renormalized) dynamics, and at the end
of the day it can happen that integrable systems just label the
classes of universality of effective actions. In other words,
the concrete type of Whitham dynamics, even if derived from the
study of integrable hierarchy, can have much broader significance.
Moreover, the Whitham equations are themselves integrable, and --
according to the previous remark -- it is mostly this integrability
that we refer to in the title of this letter.}
(these are exactly the ones that arised in the recent
studies of $2d$ topological theories/gravities \cite{DVV},
\cite{Kr2}, \cite{D}, \cite{LosPol}
and describe exact solutions to the string equations
\cite{LGGKM}, \cite{KhMar}.)

\bigskip

3. Now let us turn to the next observation.
If one takes as a characteristic of effective dynamics
in the vicinity of the classical solution the SW
formulas (\ref{spec}-\ref{lambda}), one immediately
recognizes them as familiar objects from the theory of the Whitham
equations. Namely, $\lambda$ in (\ref{aaD}) is exactly the generating
1-differential arising in the first Gurevich-Pitaevsky
problem \cite{GP}.\footnote{ This problem came from
physics of fluids and concerns the decay of a step (Heavyside)
function under the KdV evolution.  The exact KdV dynamics,
$$
\frac{\partial \tilde u}{\partial t_3} =
\tilde u\frac{\partial \tilde u}{\partial t_1}
+ \frac{\partial^3 \tilde u}{\partial t_1^3},
$$
drives the initial profile of Fig.4a into that like Fig.4b,
while the Whitham dynamics describes the smooth
enveloping curve, see Fig.4c.
For comparison, Fig.4d shows result of the evolution of the same step
function according to naive ``quasiclassical'' KdV (which is in
fact the spherical Whitham eqaution):
the Bateman-Hopf equation,
$$
\frac{\partial \tilde u}{\partial T_3} =
\tilde u\frac{\partial \tilde u}{\partial T_1}
$$
}

In formal terms, the Whitham equations
can be described as follows.
The KP/Toda-type $\tau$-function
associated with a given spectral Riemann surface is equal to

\be\label{KPsol}
{\cal T}\{t_i\} = e^{{\bf t}\gamma{\bf t}}\vartheta\left(\vec \Phi_0 +
\sum_{i=1}^{\infty}
t_i\vec k_i\right), \nn \\
\vec k_i = \oint_{\vec B} d\Omega_i(z),
\ee
where $ \vartheta$ is a Riemann theta-function and $d\Omega_i(z)$ are
meromorphic 1-differentials with poles
of the order $i+1$ at a marked point $z_0$. They are fully
specified by normalization relations

\be
\oint_{\vec A} d\Omega_i = 0
\label{norA}
\ee
and
\be
d\Omega_i(z) = \left(\xi^{-i-1} + o(\xi)\right) d\xi
\label{norc}
\ee
where $\xi$ is the local coordinate in the vicinity of $z_0$.
The moduli $\{ u_\alpha \}$ of the spectral surface are
invariants of KP flows,

\be
\frac{\partial u_\alpha}{\partial t_i} = 0,
\ee
and label the ``vacua'' -- the (finite-gap) solutions to the KP system.
The effective dynamics on the space of these ``vacua'',
generated by the Bogolyubov-Whitham method, arises with respect to
some {\it a priori} new ``slow'' Whitham times $T_i$. The way the moduli
depend on $T_i$ is defined  by the
Whitham equations (induced by the fast KP/Toda-type equations), which
for the
two-dimensional integrable systems were first derived in \cite{kr3}
in the
following form

\be
\frac{\partial d\Omega_i(z)}{\partial T_j} =
\frac{\partial d\Omega_j(z)}{\partial T_i}.
\ee
These equations imply that

\be
d\Omega_i(z) = \frac{\partial dS(z)}{\partial T_i}
\label{SS}
\ee
with some ''generating" 1-differential $dS(z)$, whose periods
can be interpreted as the effective ''slow" variables. Note that
the self-evident
relation (\ref{SS}) was crucially used in the constructing the
exact solutions
to the Whitham equation that was proposed in \cite{kr3}.
The equations for moduli, implied by this system, are of
peculiar linear form:

\be\label{whiv}
\frac{\partial u_\alpha}{\partial T_i} =
v_{ij}^{\alpha\beta}(u)\frac{\partial u_\beta}{\partial T_j}
\ee
with some (in general complicated) functions $v_{ij}^{\alpha\beta}$,
which depend on the {\it type} of ``vacua'' under
consideration
\footnote{These formulas imply a special choice of basis in the
moduli space, taking co-ordinates ( $T$-variables ) coming from
commuting KP-flows.
The relation
$\tau_{ij} = \frac{\partial^2\log{\cal T}}{\partial T_i \partial T_j}$
which defines the period matrix in terms of the
N=2 superpotential \cite{SW} has also appeared in the theory
of topological $2d$-theories, see \cite{D}}.

In the KdV case  all the spectral surfaces are
hyperelliptic, $i$ takes only odd
values $i = 2j+1$, and

\be
d\Omega_{2j+1}(z) = \frac{{\cal P}_{j+g}(z)}{y(z)}dz,
\ee
the coefficients of the polynomials ${\cal P}_j$ being fixed
by normalization conditions (\ref{norA}), (\ref{norc})
(one usualy takes $z_0 = \infty$ and the local parameter
in the vicinity of this point is $\xi = z^{-1/2}$). In this case the
equations (\ref{whiv}) can be diagonalized if coordinates
$ \left\{  u_\alpha \right\}$ on the moduli space are taken to be the
ramification points:

\begin{equation}\label{whihy}
v_{ij}^{\alpha\beta}(u) = \delta ^{\alpha\beta}\left.{d\Omega
_i(z)\over d\Omega _j(z)}\right| _{z=u_\alpha}
\end{equation}
Now an important remark is that after one swichtes on the Whitham
dynamics the periods of the differential $dS$ defined by (\ref{SS})
become the periods of the ''modulated" function (\ref{KPsol}). We will
see below that it gives us the SW spectrum.

\bigskip

4. Let us be more specific in the elliptic (GP/SW) case and restrict
ourselves to the first two time-variables, $i = 1,3$. The elliptic
(one-gap) solution to KdV is

\be
\tilde u(t_1,t_3,\ldots| u) =
\frac{\partial^2}{\partial t_1^2} \log{\cal T} (t_1,t_3,\ldots|u) = \\
= U_0 \wp (k_1t_1 + k_3t_3 + \ldots + \Phi_0 |\omega , \omega ') +
 {u\over 3}
\label{ukdvsol}
\ee
where $\wp (t)$ is the Weierstrass $\wp$-function, and

\be
dp \equiv d\Omega_1(z) = \frac{z - \alpha (u)}{y(z)}dz, \nn \\
dQ \equiv d\Omega_3(z) = \frac{z^2 - \frac{1}{2}uz - \beta (u)}{y(z)}dz.
\label{pp}
\ee
Normalization conditions (\ref{norA}) prescribe that
\be
\alpha (u) = \frac{\oint_A \frac{zdz}{y(z)}}
{\oint_A \frac{dz}{y(z)}}\ \ \ {\rm and} \ \ \
\beta (u) = \frac{\oint_A \frac{(z^2 - \frac{1}{2}uz)dz}{y(z)}}
{\oint_A \frac{dz}{y(z)}}.
\ee

The observation, that we refered to at the beginning of sect.3,
is that a particular solution $dS(z)$ to eqs.(\ref{SS}) in the elliptic case
is the same as the differential $\lambda(z)$ in
(\ref{lambda}).

Indeed, as we are going to demonstrate,

\be
dS(z) = \left(T_1 + T_3(z+\frac{1}{2}u) + o(T_5) + \ldots \right)
\frac{z-u}{y(z)}dz =
g(z|T_i,u)\lambda(z),
\label{S}
\ee
where $g(z)$ is a calculable function of Whitham times with pole
only at $z = \infty$ of the order $\frac{I-1}{2}$, if
$T_I \neq 0$ and all the $T_{i>I}=0$. The reason why $dS(z)$
has this particular form (i.e. pocesses double zero at $z=u$)
is simple.  Normally, derivative of a
meromorphic object over moduli has more poles (since after a change
of the
complex structure the holomorphic object becomes non-holomorphic),and
 moduli in
the hyperelliptic parametrization are located at ramification points.
In our
case there is just one ramification point, $u$, which is $T_i$-dependent,
and,
in order to cancel the pole at $z=u$ in $\partial dS(z)/\partial T_i$
(which
does not occur in $d\Omega_i(z)$), one needs to put some power of
$(z-u)^{1/2}$
in the numerator of $dS(z)$ -- once $y(z)$ appeared in the
denominator. Since
$(z-u)^{1/2}$ is not a single-valued function on the surface,
one needs to take
its square.

{}From (\ref{S}) one derives:
\be
\frac{\partial dS(z)}{\partial T_1} =
\left( z - u - (\frac{1}{2}T_1 + u T_3)
\frac{\partial u}{\partial T_1}\right)\frac{dz}{y(z)}, \nn \\
\frac{\partial dS(z)}{\partial T_3} =
\left( z^2 - \frac{1}{2}uz - \frac{1}{2}u^2 -
(\frac{1}{2}T_1 + u T_3)
\frac{\partial u}{\partial T_3}\right)\frac{dz}{y(z)}, \nn \\
\ldots ,
\ee
and comparison with explicit expressions (\ref{pp}) implies:
\be
(\frac{1}{2}T_1 + u T_3)\frac{\partial u}{\partial T_1} =
\alpha (u) - u, \nn \\
(\frac{1}{2}T_1 + u T_3)\frac{\partial u}{\partial T_3} =
\beta (u) - \frac{1}{2}u^2.
\label{preW}
\ee
In other words, this construction provides a (GP) solution to the
Whitham equation
\be
\frac{\partial u}{\partial T_3} = v_{31}(u)
\frac{\partial u}{\partial T_1},
\label{GPe}
\ee
with
\be
v_{31}(u) = \frac{\beta(u)-\frac{1}{2}u^2}{\alpha(u)-u} =
\left.\frac{d\Omega_3(z)}{d\Omega_1(z)}\right|_{z=u},
\ee
which can be expressed through elliptic integrals \cite{GP}.

We see that (\ref{aaD}) can be reinterpreted as

\be
\left. a = {1 \over T_1}
\oint_A dS(z)\right|_{T_3,T_5,\ldots = 0}, \nn \\
\left. a_D = {1 \over T_1}
\oint_B dS(z)\right|_{T_3,T_5,\ldots = 0}
\ee
being the periods of ''modulated" GP elliptic solution. This
implies that in generic situation (for non-elliptic surfaces
and all $T_{2j+1}\neq 0$) the SW formula (\ref{spec}) should be
\be
{\cal M}_{\vec m,\vec n} \sim \left|
\vec m \oint_{\vec A}dS + \vec n\oint_{\vec B}dS \right|
\label{specgen}
\ee
Note also that
\be
\frac{\partial}{\partial T_i}\oint_{\vec A}dS =
\oint_{\vec A} d\Omega_i = 0,
\ee
while
\be
\frac{\partial}{\partial T_i}\oint_{\vec B}dS =
\oint_{\vec B} d\Omega_i = \vec k_i,
\ee
which are the frequencies in the original KP/Toda-type solution
(\ref{KPsol}). So, the periods of the ''modulated" Whitham solution
give rise to the mass spectrum in the SW exact solution and its
generalizations.

\bigskip

5. All the quantities entering the Whitham equations have the
meaning of the averaged characteristics of the bare elliptic
solution (\ref{ell}). Note also, that the further speculation of
the meaning of the GP
solution in the SW context of $4d$ gauge theory is possible.
If one relates KP times
$ t_{1}$, $t_{3}$ with (the functions of) bare $\theta $ and
$g^{-2}$ then the main object under
consideration -- the KdV ''potential" $u(t_{1},t_{3})$ becomes
related to
the correlator $\langle F{\tilde{F}}, F{\tilde{F}}\rangle$. This
looks quite
hopefull since such correlators contain the
information about topological exitations in the gauge theory.
Now after the averaging the ''slow" times $T_{1},T_{3}$ in the
Whitham  system can be identified with the functions of
the "renormalized" KP times (coupling constants). Moreover
the form of the GP solution suggests its interpretation as
a "decay" of the topological exitations in SW theory in the
non-perturbative regime $ |u| < \Lambda $.
The GP solutions have automodel form
and this can be relatd to emergence of holomorphic
coupling constant  $\tau = \frac{1}{2\pi}(ig^{-2} + \theta)$. We will
return to this correspondence in a separate publication.

\bigskip

6. To conclude, we see that the central formula
(\ref{spec}) of \cite{SW} can be interpreted as
(\ref{specgen}), i.e. in terms of
periods of the central object $dS(z)$ in the theory of the Whitham
hierarchy. This observation seems to be important since there
exists a general beleif that low-energy effective actions are
proper objects to be reffered to as {\it generalized}
$\tau $-functions.
One should add that conceptually the
Whitham method is precisely the averaging over fast fluctuations,
which is necessary to produce the effective action for slow
variables i.e. plays the role of the non-perturbative analog of
renormalization group.
We believe that this analogy deserves attention and further
studies\footnote{Among related studies, one should mention
\cite{VW}, \cite{LMNS}, \cite{DS} and \cite{KS}.
It also deserves noting that -- as usual -- the study of
some $2d$ analogues of the $4d$ YM theory can shed an additional
light on the problem. The obvious relation is to
$2d$ topological theories and string equations. In both
cases the Whitham dynamics is known to arise in
description of effective actions (see \cite{GM} for discussion of
parallels with the $2d$ physics).}
will put them on a more solid ground.

\bigskip

We are indebted to E.Akhmedov, V.Fock, A.Gurevich, S.Kharchev,
I.Polyubin, V.Rubtsov,
K.Selivanov, A.Smilga and A.Zabrodin for
fruitfull discussions. We aknowledge the hospitality of A.Niemi
at Uppsala University where part of this work was done. A.G  thank
to H.Leutwyler for the  warm hospitality in the Bern University
where this work was finished.
The work of A.G., A.Mar. and A.Mir. is partly supported by
 ISF grant MGK000 and grants of Russian fund of fundamental research
RFFI 93-02-03379 and RFFI 93-02-14365.
The work of A.Mar. was also supported by NFR-grant No F-GF 06821-305
of the Swedish Natural Science Research Council.

\newpage
$$
{\epsfxsize 7cm\epsfbox{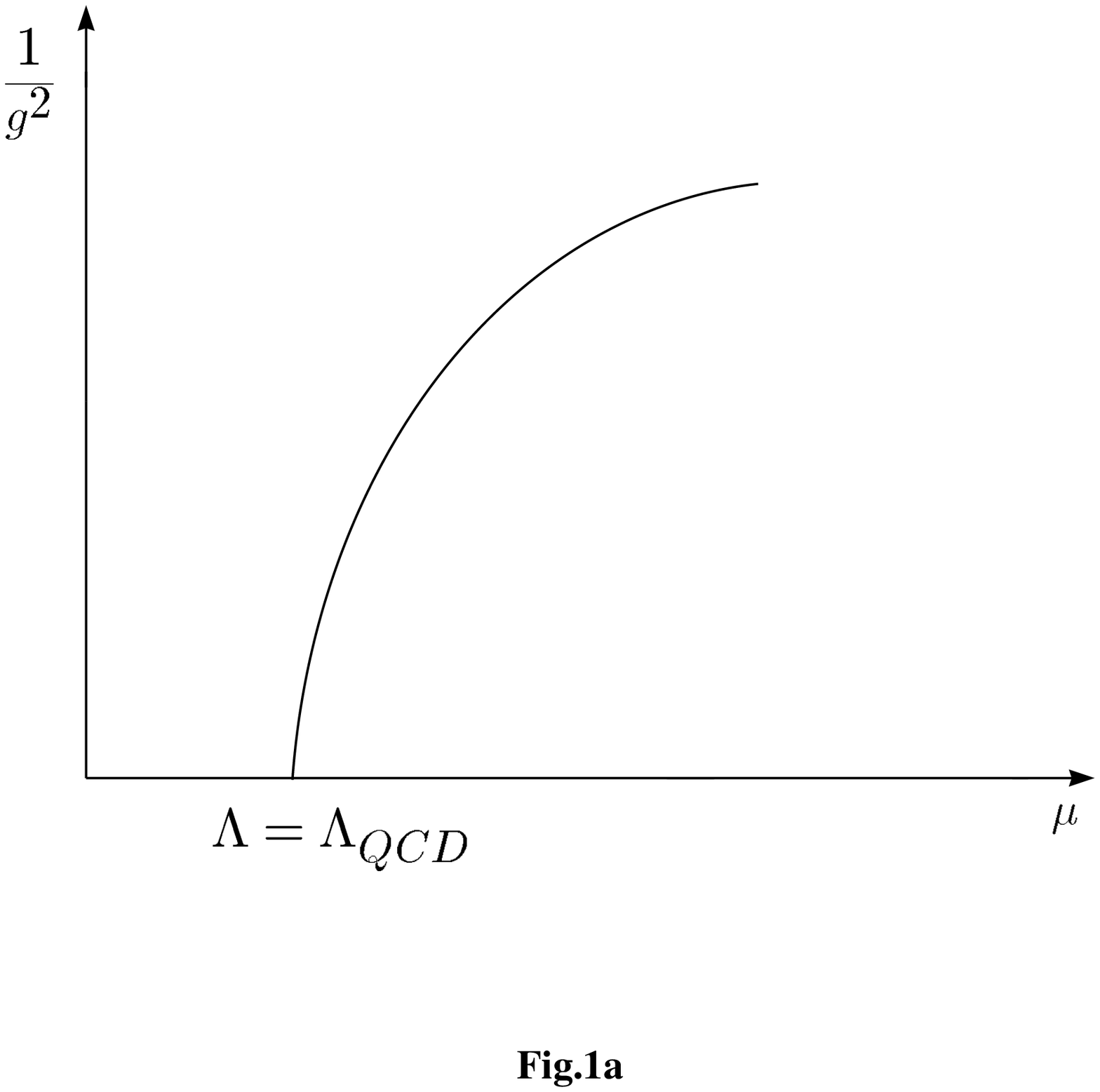}}
$$

$$
{\epsfxsize 8.5cm\epsfbox{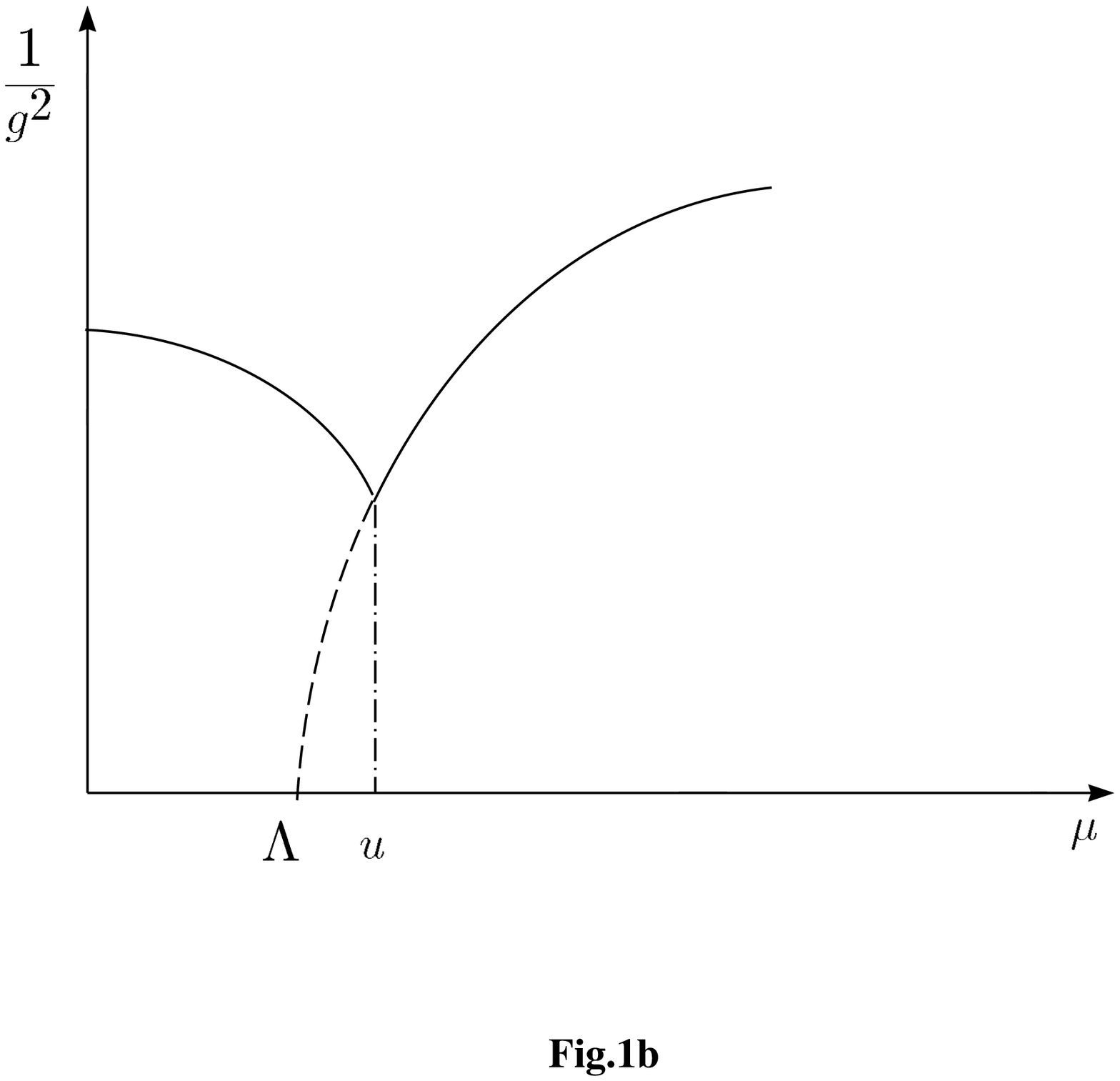}}
$$

$$
{\epsfxsize 7cm\epsfbox{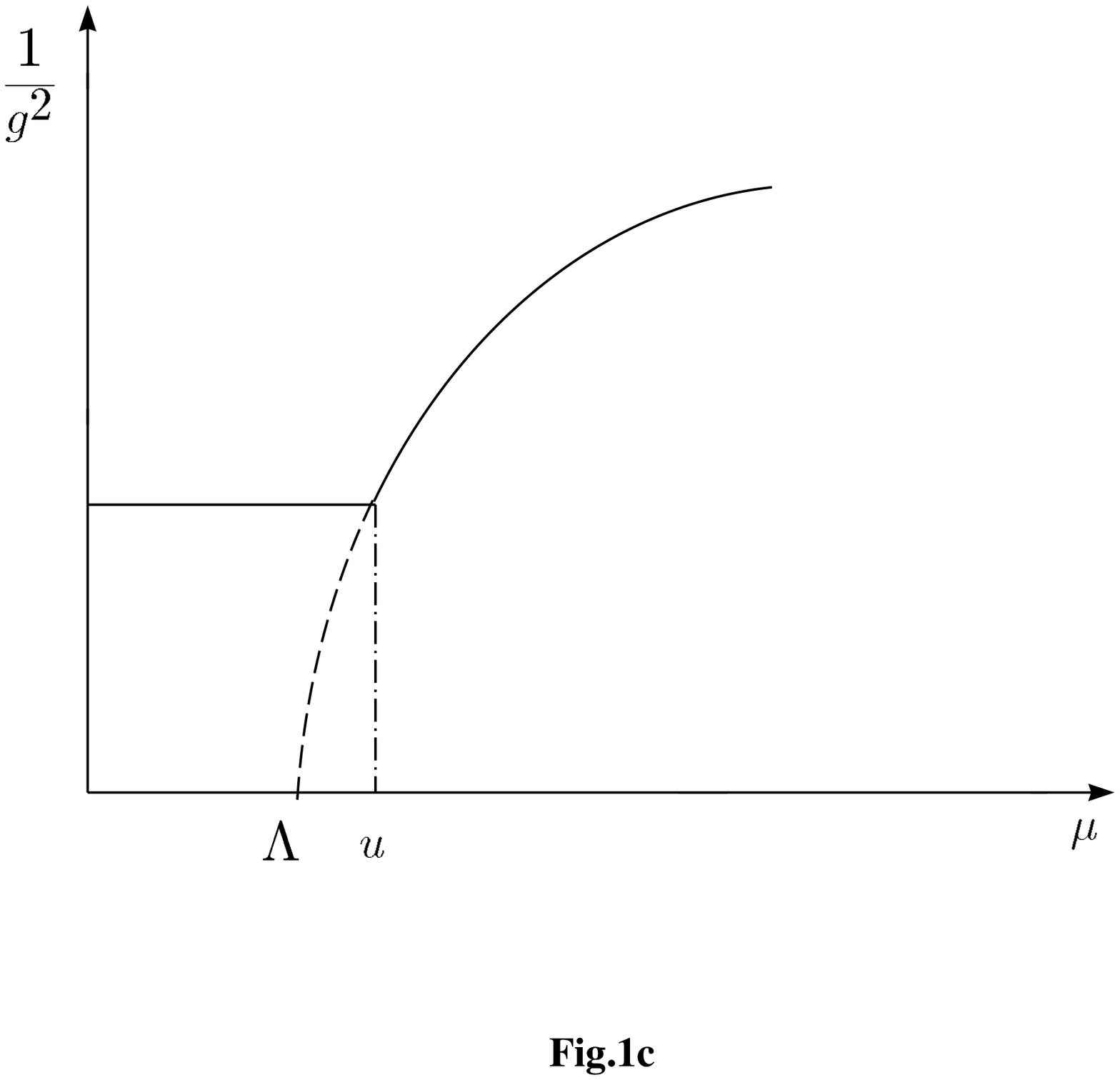}}
$$

$$
{\epsfxsize 7cm\epsfbox{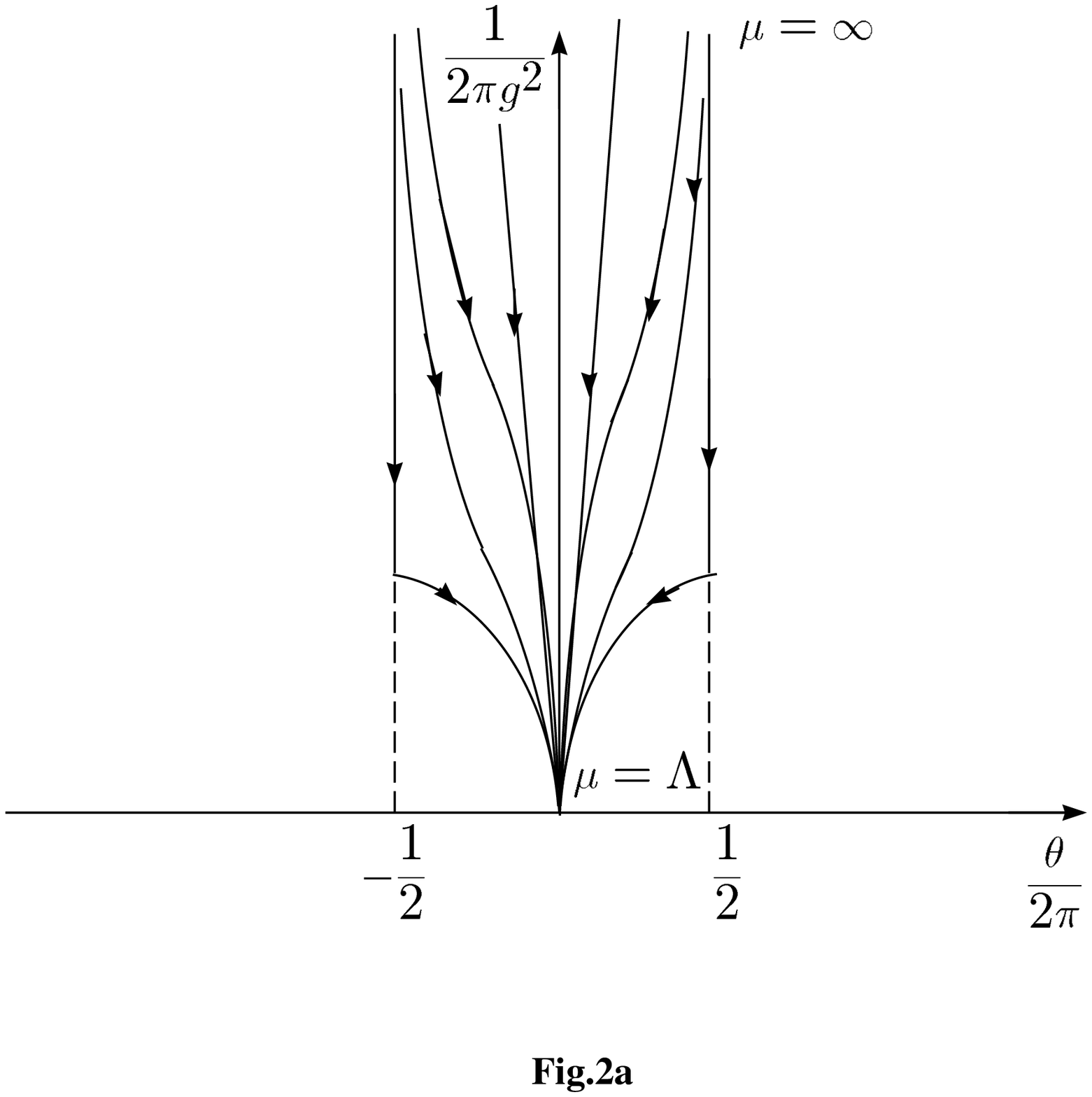}}
$$

$$
{\epsfxsize 7cm\epsfbox{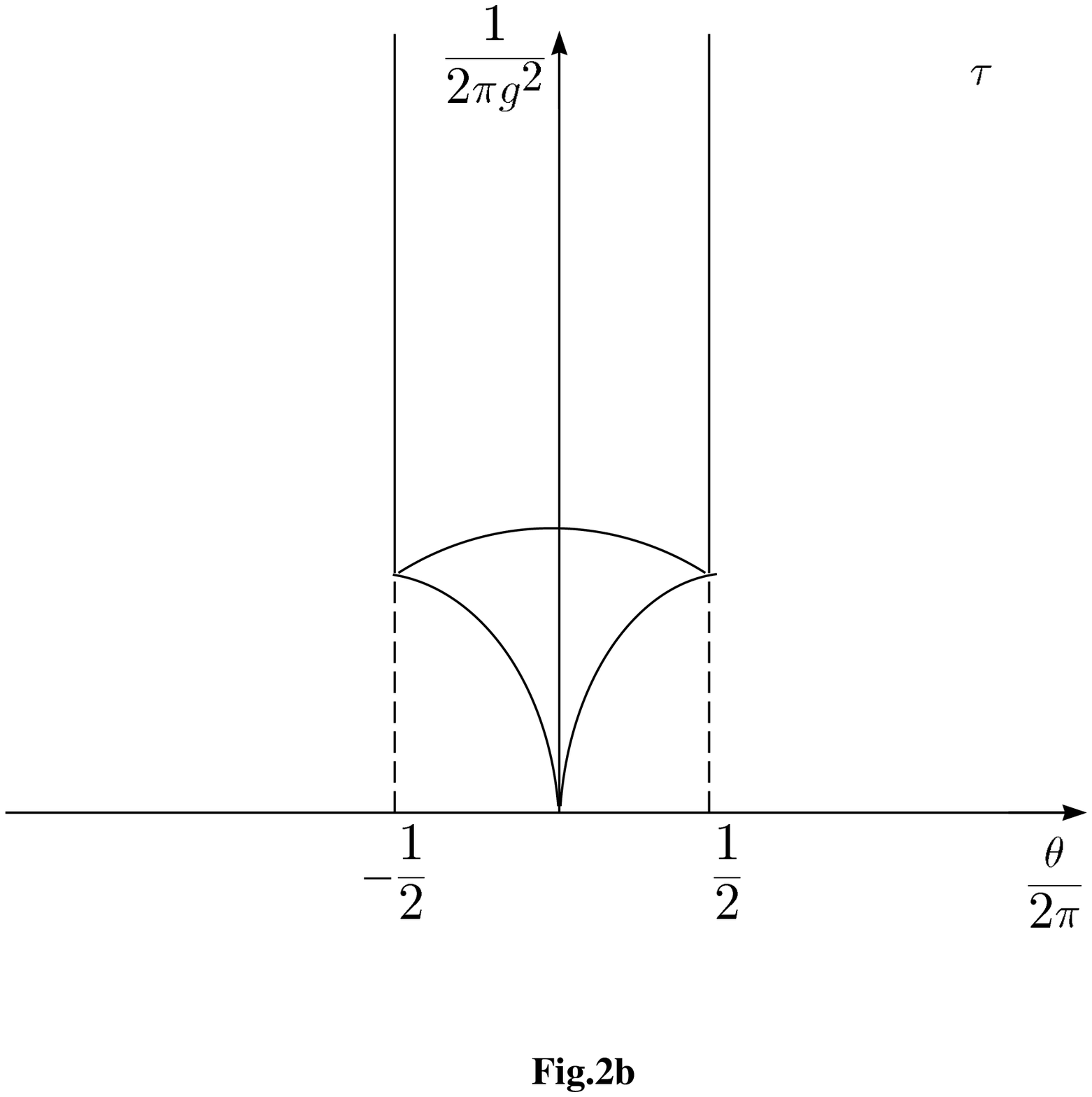}}
$$

$$
{\epsfxsize 7cm\epsfbox{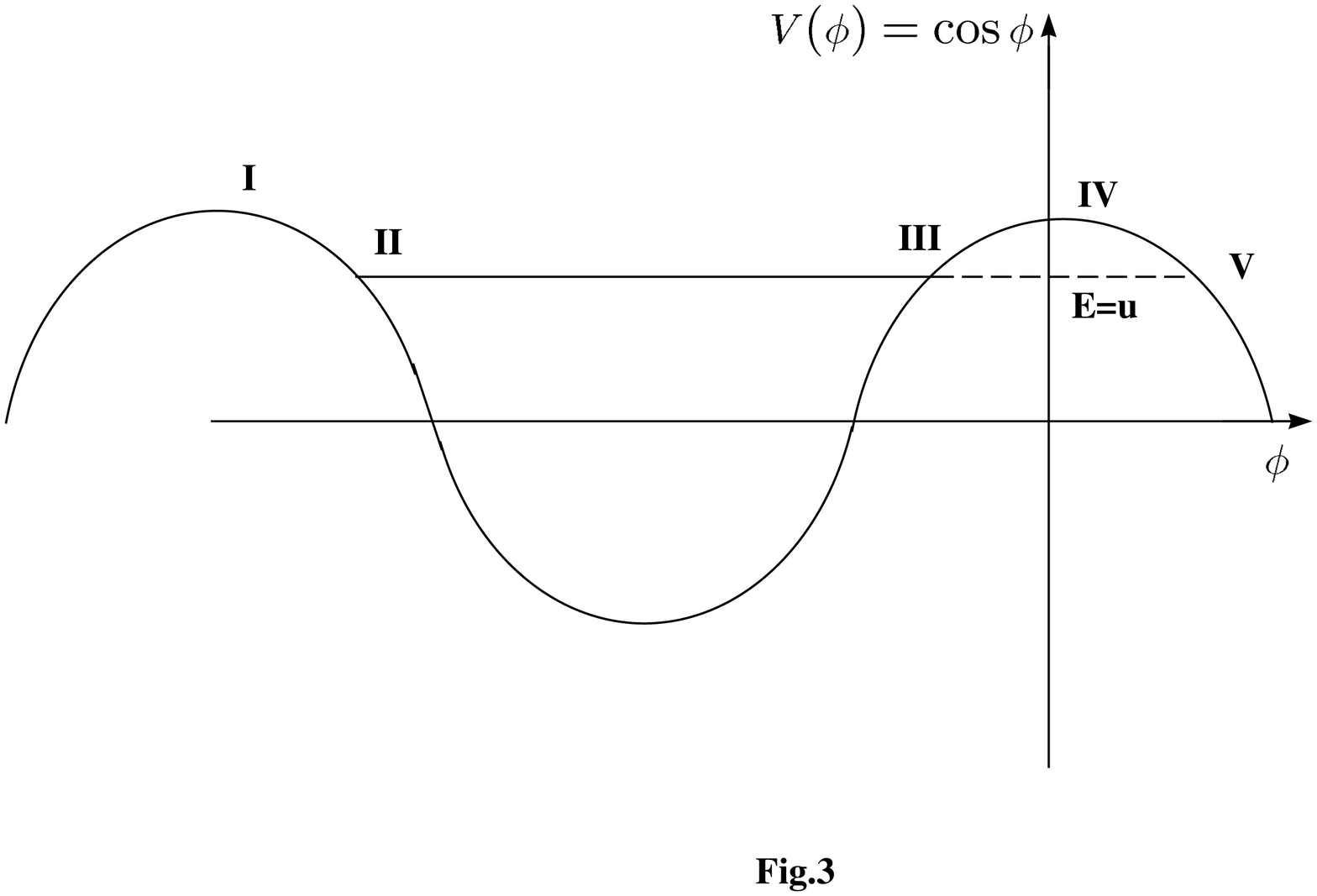}}
$$

$$
{\epsfxsize 7cm\epsfbox{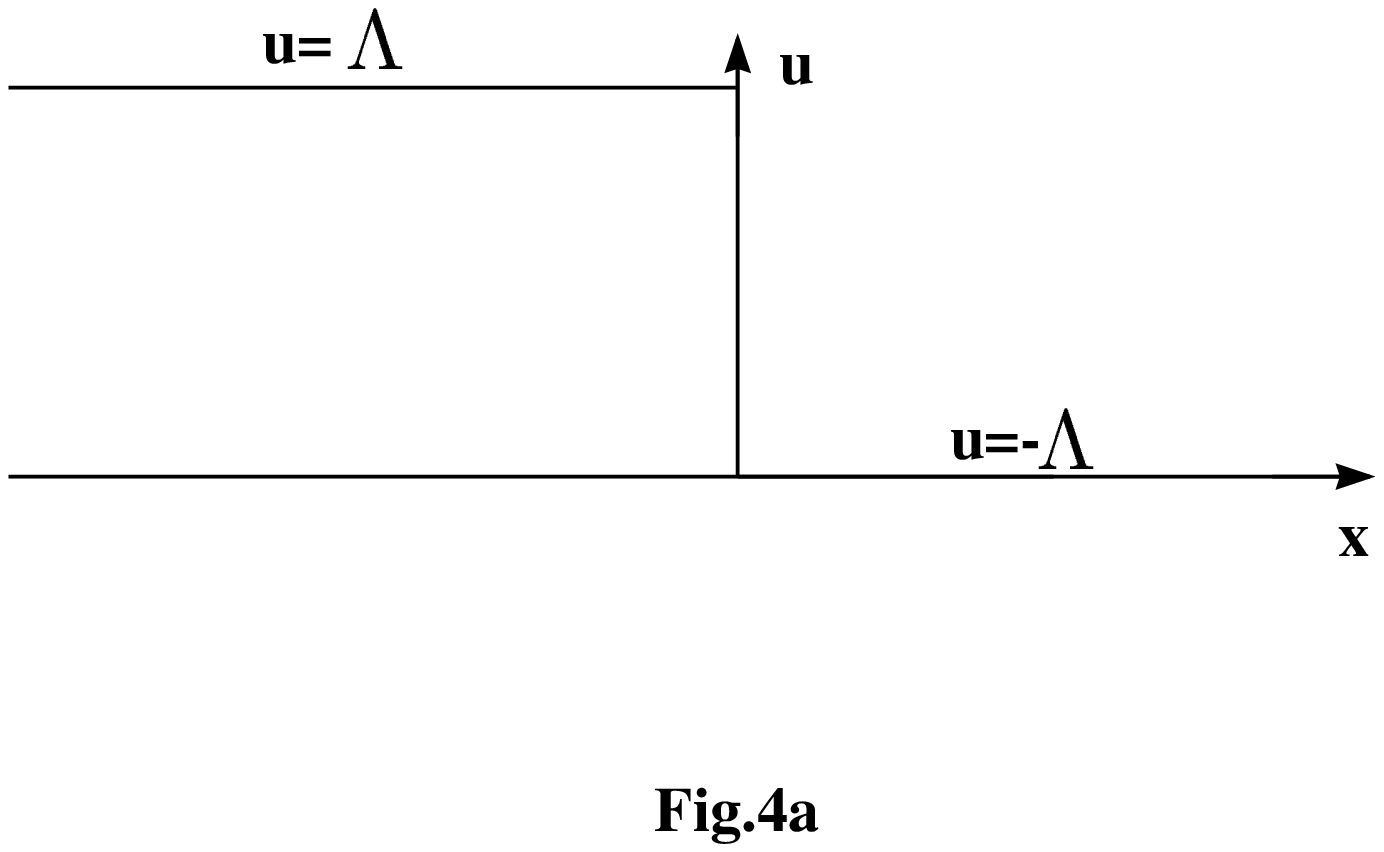}}
$$
$$
{\epsfxsize 7cm\epsfbox{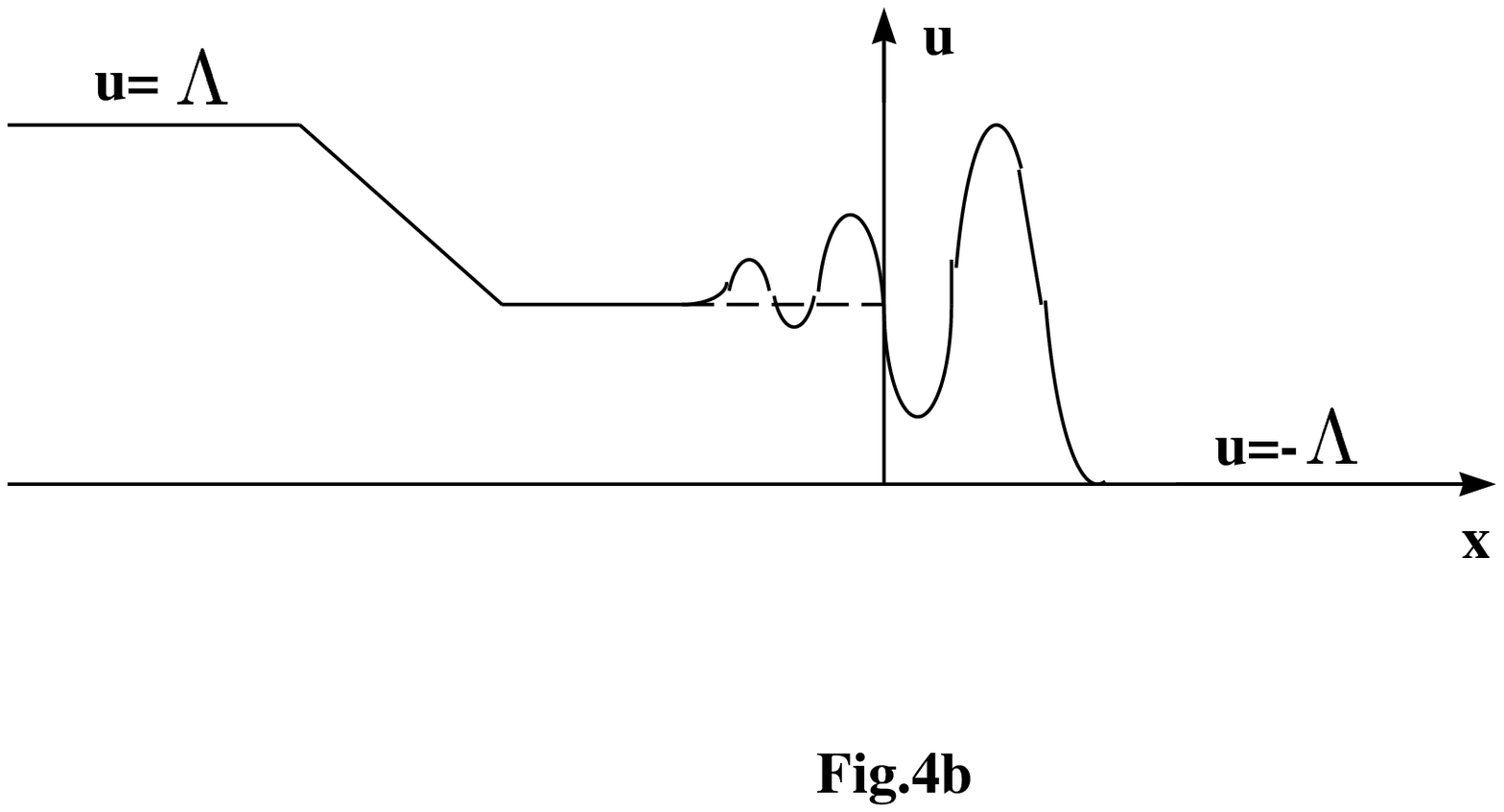}}
$$
\vspace{10cm}
$$
\begin{picture}(50,50)(50,30)
\put(0,13){\makebox(0,0){{\epsfxsize 11.5cm\epsfbox{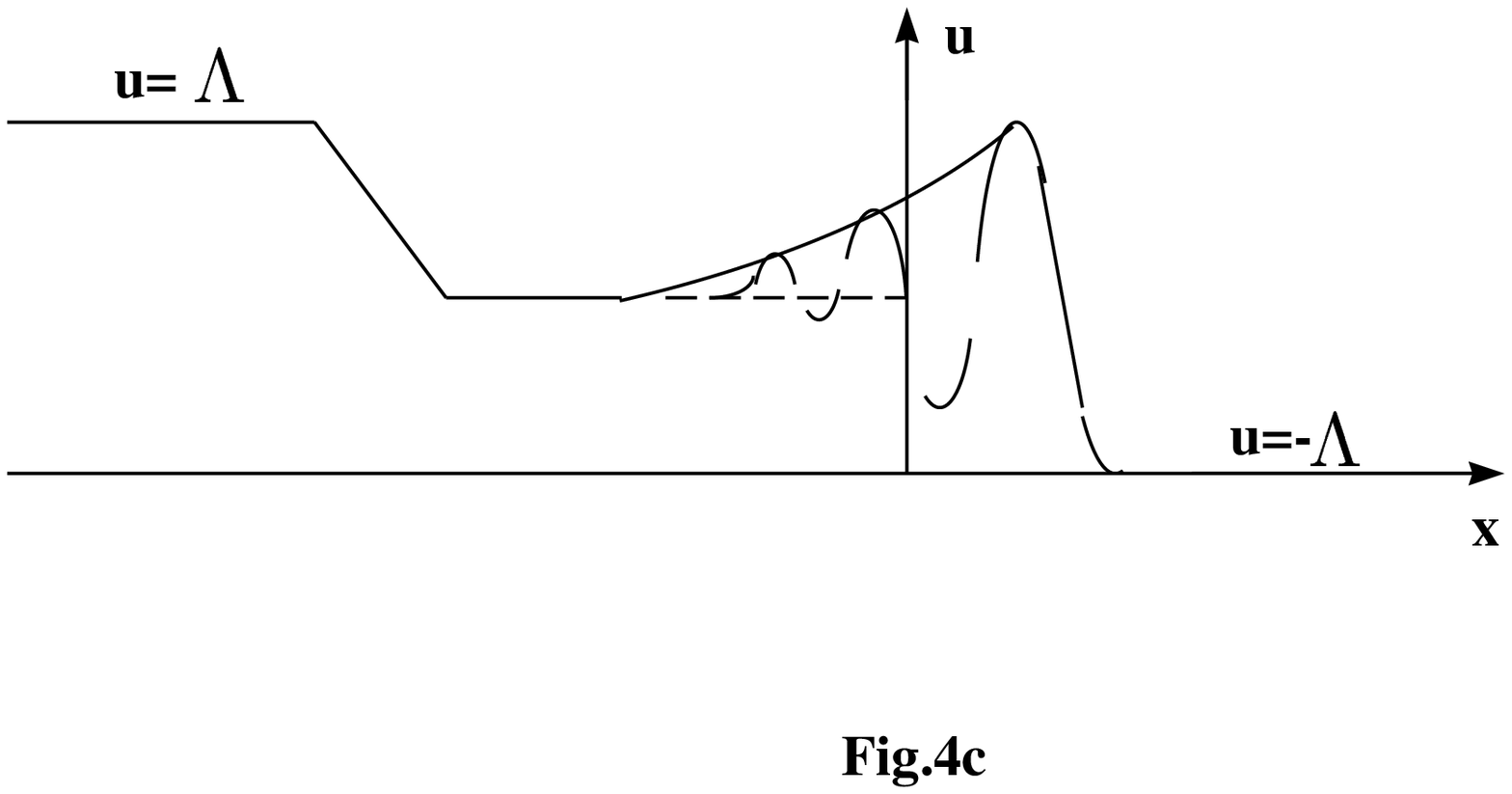}}}}
\end{picture}
$$
\vspace{4.5cm}
$$
{\epsfxsize 7cm\epsfbox{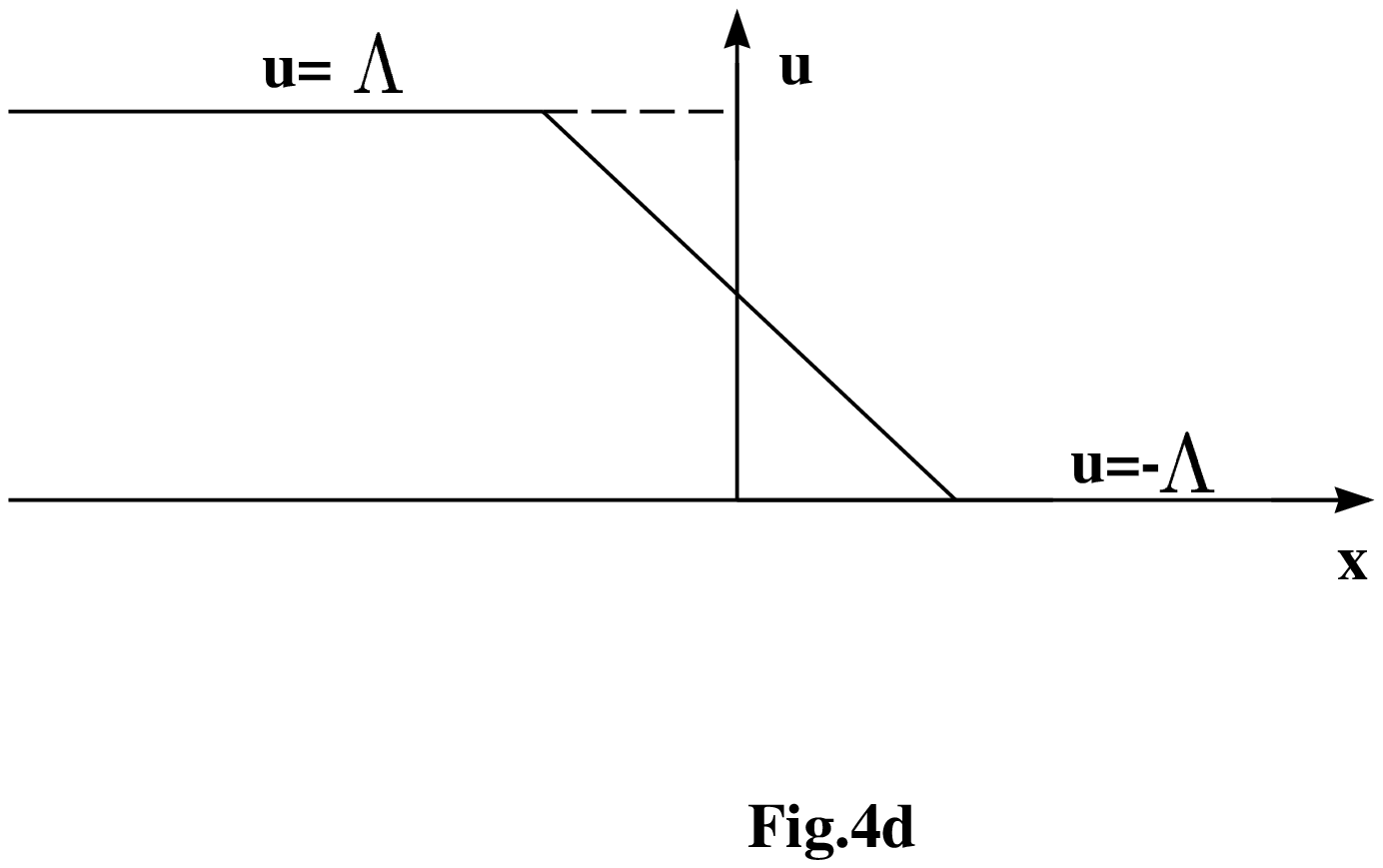}}
$$

\end{document}